\documentclass{elsart}
\usepackage{psfig}

\begin{document}
\begin{frontmatter}

\title{Exploring the Time Domain with the Palomar-QUEST Sky Survey}

\author[caltech]{A. Mahabal},
\author[caltech]{S. G. Djorgovski},
\author[caltech]{M. Graham},
\author[caltech]{R. Williams},
\author[caltech]{B. Granett},
\author[caltech]{M. Bogosavljevic},
\author[yale]{C. Baltay},
\author[yale]{D. Rabinowitz},
\author[yale]{A. Bauer},
\author[yale]{P. Andrews},
\author[yale]{N. Morgan},
\author[yale]{J. Snyder},
\author[yale]{N. Ellman},
\author[yale]{S. Duffau},
\author[indiana]{J. Musser},
\author[indiana]{S. Mufson},
\author[indiana]{M. Gebhard},
\author[ncsa]{R. Brunner} and
\author[ncsa]{A. Rengstorf}
\address[caltech]{California Institute of Technology, Pasadena, CA 91125}
\address[yale]{Yale University, New Haven, CT 06520}
\address[indiana]{Indiana University, Bloomington, IN 47405}
\address[ncsa]{NCSA/UIUC, Champaign, IL 61820}

\end{frontmatter}
\section{Introduction}
Discoveries in astronomy are often made through a systematic exploration of
previously poorly covered regions of the observable parameter space
\cite{djorg01a,djorg01b}.
In particular, exploration of the time variability on the sky over a broad
range of flux levels and wavelengths is rapidly becoming a new frontier
of astronomical research
\cite{paczynski,diercks}.
Many exciting astrophysical phenomena are known in the time domain:
all manner of variable stars,
stellar explosions such as SNe and GRBs,
variable AGN, pulsars, microlensing events,  etc.
Yet undoubtedly many more remain to be discovered.

A number of surveys and experiments exploring the time domain are already
in progress 
(see, e.g., \cite{pacz01,paczweb} for listings),
and even more ambitious synoptic sky surveys are being planned, e.g.,
\cite{tyson,kaiser}.
The important factors
in such programs are (1) the area covered, (2) the depth of coverage, (3)
number of wavelengths used, and (4) the baseline(s) in time. 

We conducted an exploratory search for highly variable objects \cite{granett}
and optical transients \cite{mahabal} 
using 
$\sim 8000~deg^2$ in the NGP and SGP areas of the 
Digital Palomar Observatory Sky Survey (DPOSS) plate overlap regions.
The effective depth of these searches was $r_{max} \approx 19$ mag for the
``high'' states, with a plate limits $r_{max} \approx 21$ mag.
Time baselines ranged from days to years, with $\sim 2$ yrs being
typical.
After eliminating various artifacts and contaminants, and applying well
defined statistical criteria for selection, we identified a large number of
highly variable objects, and followed up spectroscopically a subset of them
at the Palomar 200-inch telescope.  They turned out to be a heterogeneous
collection of flaring M-dwarfs, OVV QSOs and BL Lacs, CVs (including a rare magnetic one), and some otherwise non-descript stars.  Approximately a 
third to a half
of all highly variable objects down to this magnitude level, at moderate and
high Galactic latitudes appear to be associated with AGN.

We also found a number of optical transients (operationally defined as
high-S/N, PSF-like objects, detected only once).
We estimate that a single-epoch ``snapshot'' down to this flux level contains
up to $\sim 1000$ transients/sky.
Their nature remains unknown, but in at least 2 cases deep follow-up imaging
revealed apparent faint host galaxies, which now await spectroscopy.

This pilot study gave us some hints as to what may be expected in a dedicated,
wide-field, synoptic sky survey at comparable magnitudes.  The faint variable
sky has a very rich and diverse phenomenology.

In this paper, we describe briefly a new, large, digital synoptic sky survey,
Palomar-QUEST (hereafter {\it PQ}); a more detailed description of the
survey will be presented elsewhere.

\begin{figure}
\centerline{
\psfig{file=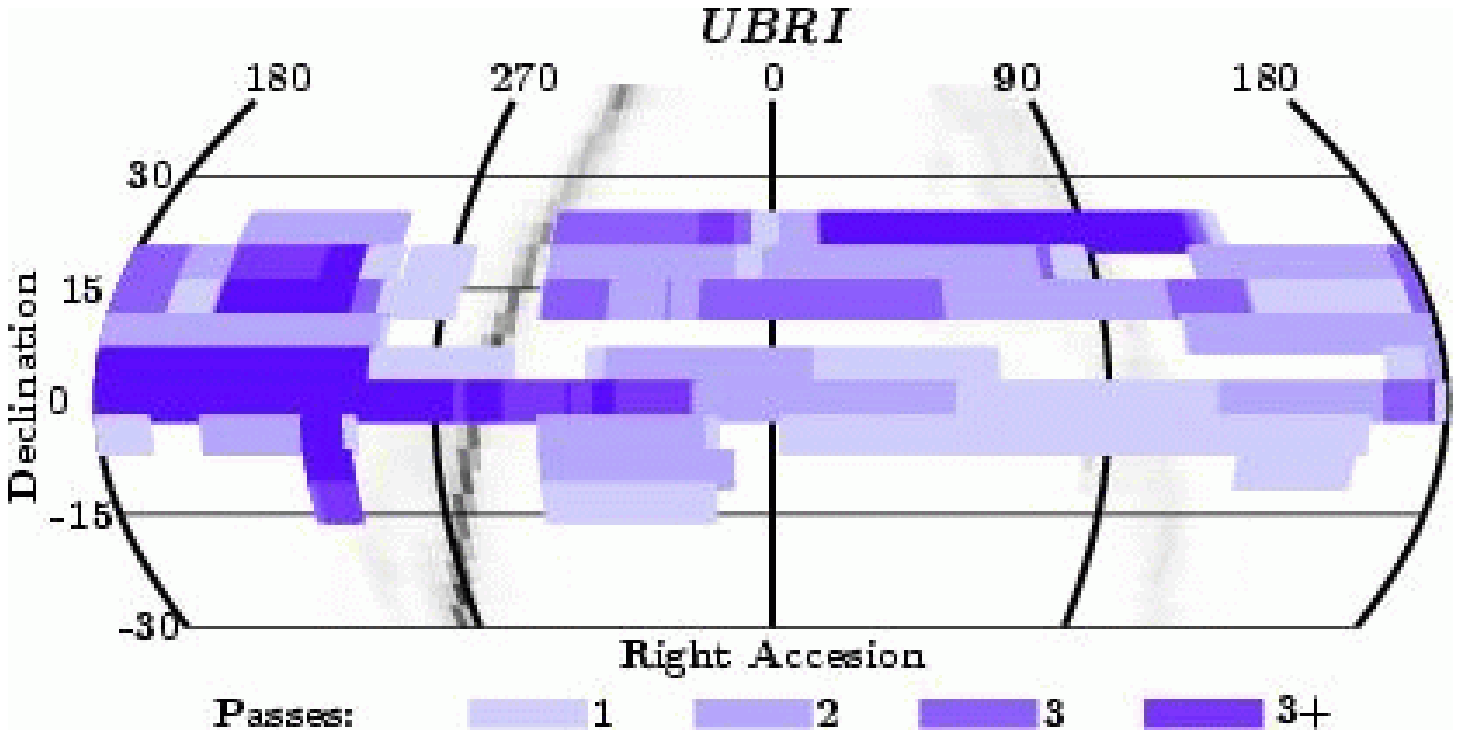,width=2.8in,angle=0}
\psfig{file=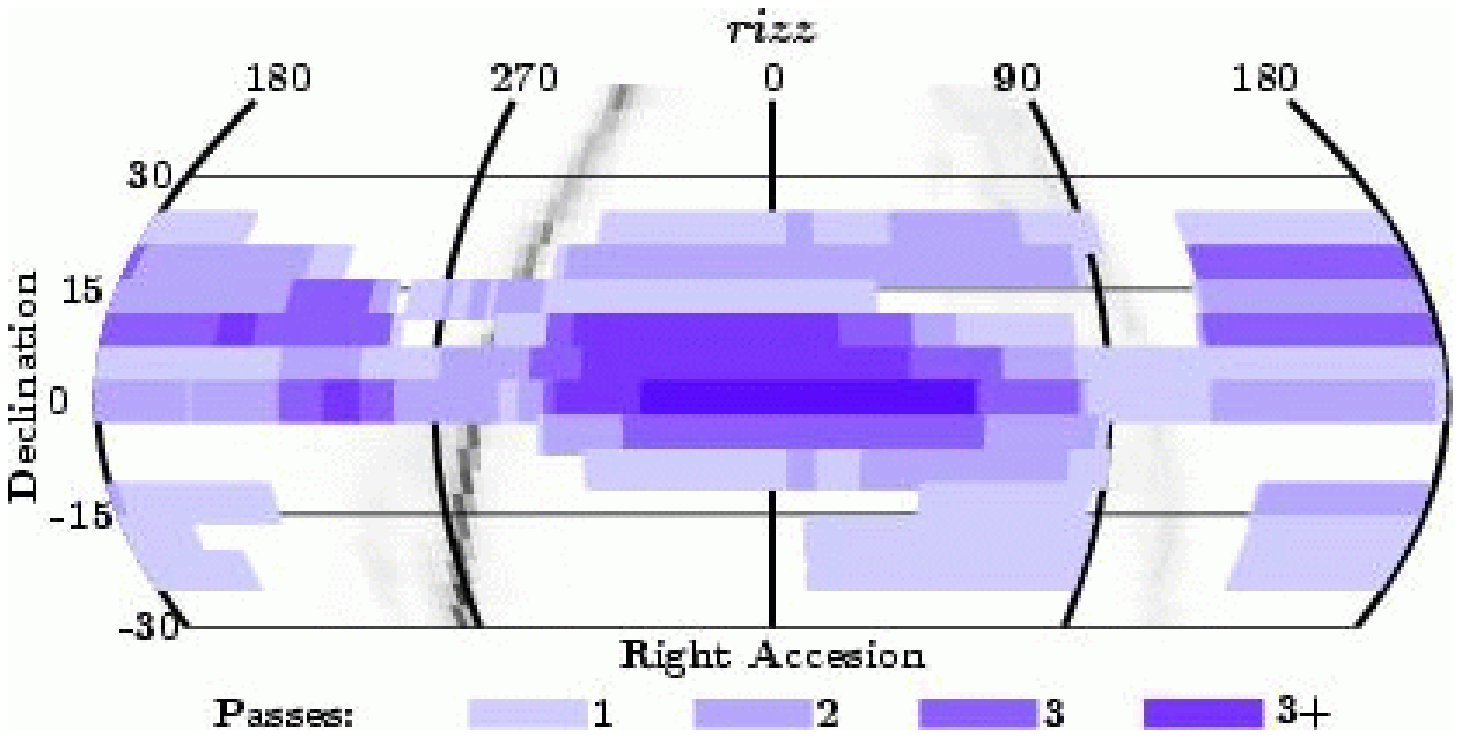,width=2.8in,angle=0}
}
\caption{
Palomar-QUEST coverage as of June 2004:  
$\sim5000 ~deg ^2$ have been covered at least twice
in {\it UBRI} and $\sim 5500~deg^2$ at least twice in {\it r'i'z'z'}.
}
\end{figure}

\section{The Palomar-QUEST Sky Survey}
The Palomar-Quest synoptic sky survey,
a collaborative project between Yale, Caltech,
and NCSA (some other groups are also involved in more specific collaborations)
is a new major digital sky survey conducted at the Samuel Oschin 48-inch 
Schmidt telescope at Palomar.  
The survey uses a special 112-CCD, 162-Megapixel camera built especially for
this purpose.  Some of the salient features of the survey are:
(1) Data taking in the Point-and Stare (PS) mode, covering $\sim 9.2 ~deg^2$
per exposure, or in a Drift Scan (DS) mode, in strips $4.6^{\rm o}$ wide, with a
typical coverage of $\sim 500 ~deg^2$/night,
(2) Near simultaneous observations in one of two filter sets in the DS mode:
Johnson-Cousin's {\it UBRI} or SDSS {\it r'i'z'z'},
(3) In good conditions, typical limiting magnitudes for point sources:
$R_{lim} \approx 22$ mag, $I_{lim} \approx 21$ mag,
(4) In the DS mode, useful Declination range 
$-25^{\rm o} < \delta < +30^{\rm o}$, for a
total anticipated survey area of $\sim 12,000 - 15,000 ~deg^2$,
(5) Multiple-pass coverage, with at least 4 passes per year at each covered
location,
(6) Time baselines for repeats ranging from days to months, anticipated to
extend to multi-year time scales over the next 3 to 5 years or beyond,
(7) NVO standard, protocols, and connections built in from the start.

The survey has started producing a steady stream of science-grade data, 
from summer of 2003.  In the DS mode, it typically generates $\sim 1$ TB of
raw image data per month (assuming $\sim 14$ clear nights).
This unprecedented amount of data makes this the largest synoptic survey of its
kind both in terms of area covered and depth.  A broad range of science is
envisioned for the survey, but exploration of the time domain will be one of
the main focal areas.
\section{Toward Real-Time Transient Detections}
The existing {\it PQ} pipeline is geared to complete processing of a night's data  by the next day.
In a matter of hours catalogs become available in the four filters used and can
be integrated and combined with other epochs available for the area covered
that night.
This is sufficient for most {\it PQ} projects, including those involving
variable objects, e.g. SNe.  However, such a turn-around time can be too slow
for the follow-up of rapidly fading sources and transients.  Thus, we have
started work on a real-time processing pipeline which will enable detections
of such sources within minutes or less.

We will be comparing nightly catalogs with older catalogs from {\it PQ} itself,
as well as other surveys and archives, using NVO infrastructure and methodology.
Positions of possible transients will be compared with those of known variable sources, known asteroids, etc. 
Our goal is to provide detections of potentially exciting sources in real-time,
via email alerts and a dedicated website.
Given the large area coverage of 
{\it PQ} (Figure 1 shows coverage current as of June 2004)
and the results from our exploratory DPOSS project noted above, we estimate
that we will be detecting up to several tens of highly variable or transient
sources per night.
A key challenge will be to deal with this abundance of data in an effective
manner -- maintaining a high completeness in terms of the interesting variable
and transient sources discovered, while maintaining a low contamination rate
by spurious or uninteresting sources -- and doing it in real time.
A number of advanced statistical and Machine Learning techniques will be
explored to this end.

\begin{figure}
\centerline{\psfig{file=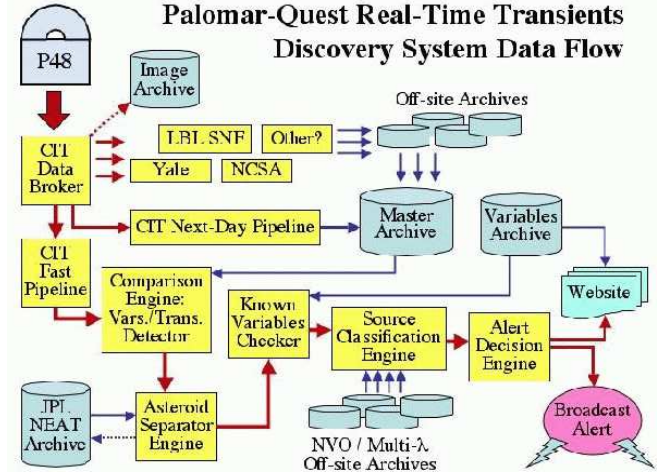,width=3.4in,angle=270}}
\caption{
A schematic outline of a real-time data reduction pipeline which
would enable a rapid discovery and spectroscopic follow-up of transients and
other interesting objects. Particular emphasis will be on eliminating spurious
or relatively uninteresting candidates, in order to keep the number of
real-time alerts reasonable, while not missing any interesting ones.
The actual design of this system is still in progress.
}
\end{figure}

We believe that {\it PQ} will provide a major new window into the faint
variable sky at optical wavelengths over the next several years.  At the same
time, this survey will serve as a science, technology, and methodology
precursor for the more ambitious future projects such as LSST \cite{tyson}
or Pan-STARRS \cite{kaiser}.
{}
\end{document}